\documentclass[10pt,letterpaper]{article}
\usepackage[top=0.85in,left=2.75in,footskip=0.75in]{geometry}
\usepackage{algorithm}
\usepackage[noend]{algpseudocode}

\usepackage{amsmath,amssymb}

\usepackage{changepage}

\usepackage[utf8x]{inputenc}

\usepackage{textcomp,marvosym}

\usepackage{cite}

\usepackage{nameref,hyperref}

\usepackage[right]{lineno}

\usepackage{microtype}
\DisableLigatures[f]{encoding = *, family = * }

\usepackage[table]{xcolor}

\usepackage{array}

\newcolumntype{+}{!{\vrule width 2pt}}

\newlength\savedwidth

\raggedright
\usepackage[aboveskip=1pt,labelfont=bf,labelsep=period,justification=raggedright,singlelinecheck=off]{caption}

\makeatletter
\renewcommand{\@biblabel}[1]{\quad#1.}
\makeatother

\usepackage{lastpage,fancyhdr,graphicx}
\usepackage{epstopdf}
\fancyheadoffset[L]{2.25in}
\fancyfootoffset[L]{2.25in}
\newcommand\reviseold[1]{\textcolor{black}{#1}}
\newcommand\revise[1]{\textcolor{black}{#1}}
\begin{document}

%\vspace*{0.2in}

\begin{flushleft}
{\Large
\textbf\newline{\reviseold{Complexity and Diversity in}  Sparse Code Priors Improve Receptive Field Characterization of Macaque V1 Neurons} 
}

\vspace*{0.2in}

Ziniu Wu\textsuperscript{1,2},
Harold Rockwell\textsuperscript{1,3,+},
Yimeng Zhang \textsuperscript{1,4},
Shiming Tang\textsuperscript{5},
Tai Sing Lee\textsuperscript{1,4, *}

\bigskip
\textbf{1} Center for the Neural Basis of Cognition and Neuroscience Institute, Carnegie Mellon University, Pittsburgh, PA, USA.
\\
\textbf{2} Department of Mathematics, Carnegie Mellon University, Pittsburgh, PA, USA.
\\
\textbf{3} Department of Biological Sciences, Carnegie Mellon University, Pittsburgh, PA, USA.
\\
\textbf{4} Computer Science Department, Carnegie Mellon University, Pittsburgh, PA, USA.
\\
\textbf{5} Center for Life Sciences, Peking University, Beijing, China.
\bigskip

* corresponding author: Tai Sing Lee,  taislee@andrew.cmu.edu \\
+ co-first author

\vspace*{0.2in}

Current Affiliation: Ziniu Wu, Department of EECS,  Massachusetts Institute of Technology; Harold Rockwell, Department of Neuroscience, University of Chicago. 

\end{flushleft}

\clearpage

% Please keep the Author Summary between 150 and 200 words
% Use first person. PLOS ONE authors please skip this step. 
% Author Summary not valid for PLOS ONE submissions.   
%\section*{Author summary}

\nolinenumbers

\section*{Abstract}

System identification techniques---projection pursuit regression models (PPR\reviseold{s}) and convolutional neural networks (CNNs)---provide state-of-the-art performance in predicting visual cortical neurons' responses to arbitrary input stimuli. 
\reviseold{However, the constituent kernels recovered by these methods are often noisy and lack coherent structure, making it difficult to understand the underlying component features of a neuron's receptive field.} 
In this paper, we show that using \reviseold{a dictionary of diverse kernels with complex shapes} learned from natural scenes based on efficient coding theory, as the front-end for PPR\reviseold{s} and CNNs can improve their performance in neuronal response prediction \reviseold{ as well as algorithmic data efficiency and convergence speed}.  Extensive experimental results also indicate that these sparse-code kernels provide important information on the component features of a neuron's receptive field. In addition, we find that models with the \reviseold{complex-shaped} sparse code front-end are significantly better than models with a standard orientation-selective Gabor filter front-end for modeling V1 neurons that have been found to exhibit complex pattern selectivity. \reviseold{We show that the relative performance difference due to these two front-ends can be used to produce a sensitive metric for detecting  complex selectivity in V1 neurons.}

\section*{Author Summary}
Convolution neural networks and projection pursuit regression models are two state-of-the-art approaches to characterizing the neural codes or the receptive fields of neurons in the visual system. However, the constituent kernels recovered by these methods are often noisy and difficult to interpret. Here, we propose an improvement of these standard methods by using a set of neural codes learned from natural scene images based on the convolutional sparse coding theory as priors or the front-end for these methods. \reviseold{We found that this approach improves the model performance in predicting neural responses with less data and with faster convergence for fitting}, \reviseold{and allows a possible interpretation of the constituents of the receptive fields in terms of the dictionary learned from natural scenes.}  \reviseold{The relative performance difference due to these two front-ends has been shown to produce an effective metric for detecting  complex selectivity in V1 neurons.} 

\section*{Introduction}

The neural code of neurons in the primary visual cortex has been investigated for decades~\cite{10, 11, 12}. A number of quantitative approaches have been developed to characterize and model the receptive fields (RFs) of V1 neurons, notably energy models~\cite{1}, spike-triggered average~\cite{2, Ringach2001} and spike-triggered covariance methods~\cite{3, 4}, linear-nonlinear cascades~\cite{5}, sub-unit models~\cite{6} and general linear models based on handcrafted nonlinear feature spaces~\cite{7, 8, 9}. The typical neuronal RFs assumed or recovered by the previous methods are Gabor filter models followed by a threshold or quadratic nonlinearity~\cite{10,11,12}, for simple and complex cells respectively. However, a recent study~\cite{tang} discovered that a large proportion of neurons in the superficial layer of V1 of awake macaque monkeys are highly selective to specific complex features~\cite{4, 16}, suggesting that many V1 neurons act also as complex pattern detectors rather than just Gabor-based oriented edge detectors. \reviseold{This finding underscores the traditional models' limitations in capturing the full characteristics of neural tuning}.

\smallskip

Sparse coding or efficient coding theory~\cite{18, 28, olshausenovercomplete, CSC, gregor2010learning} has provided great insight into the computational principles underlying the structure of the receptive fields of V1 neurons. 
\reviseold{In the highly overcomplete regime~\cite{olshausen2013highly}, particularly when coupled with a convolutional mechanism~\cite{CSC}}, sparse coding yields a diverse set of features, including center-surround filters, corner, curvature, and cross detectors.
In this paper, we investigate the idea that a set of basis filters learned from natural scene images based on convolutional sparse coding could be used to improve PPR\reviseold{s} and CNNs in neuronal RF recovery tasks. \reviseold{In short}, the basic idea is to use these learned codes as the front-end for PPR\reviseold{s} and CNNs to predict the neuronal responses. Our experiments show that CNNs with a diverse \reviseold{complex-shape} code front-end not only achieve the state-of-the-art performance in neuronal response prediction, but are also more data-efficient to train and \reviseold{faster in convergence}. This result suggests that  \reviseold{this more diverse set of } basis functions learned from natural scenes might provide a better approximation of the underlying components of the neurons' RFs than the \reviseold{noisy} filters \reviseold{, characterized by white-noise like power spectrum and lacking in coherent structures, often} learned by CNN trained from scratch.

Earlier studies showed that sparse coding principle leads to the development of Gabor-like receptive fields~\cite{16, 18}, whereas recent work suggested that receptive fields or kernels with more complex feature selectivity would arise when there are more neurons (basis functions) available than necessary, i.e. overcomplete representation, \revise{obtained} either with more features or denser spatial sampling (\revise{due to} convolution) ~\cite{18, olshausenovercomplete, 28, CSC, 27}. These complex features and codes are consistent with the diversity and the complexity of the feature tuning observed in V1~\cite{tang, Victor2006, Livingstone2017, Dobbins1987, Sillito1995, Ringach2002}. Therefore, we conjecture that neurons selective to higher-order complex features, such as corners, curvatures, and junctions, might be better modeled by the \reviseold{complex-shaped kernels provided by convolutional sparse coding} \revise{than by a less diverse set of kernels.} To test this conjecture, we compare the performance of CNN models with two different front-end types (\reviseold{complex-shaped} codes versus Gabor wavelets). Indeed, we found that models with the complex front-end are much better than models with the classical Gabor and/or Laplacian front-end for modeling and predicting the activities of V1 neurons, and this improvement is larger for neurons previously determined to have complex pattern selectivity. This observation 
further supports \reviseold{ a plausible role} of the convolutional sparse coding theory for understanding neural codes in V1.

\smallskip
\noindent\underline{\textbf{Contributions:}}
\reviseold{We introduce two novel classes of neural response prediction models \reviseold{for receptive field characterization}:(1) convolution matching pursuit regression models (CMPRs), a variant of PPRs, which select the kernel from the sparse coding dictionaries instead of learning from scratch; 
(2) fixed-kernel CNNs (FKCNNs), which are CNNs with fixed first-layer kernels.  FKCNNs  achieves state-of-the-art performance with improved data-efficiency \reviseold{and faster convergence.}}
\reviseold{Both methods allow possible interpretation of the constituents of the receptive fields in terms of the dictionary learned from natural scenes, and provide  useful metrics for characterizing and discriminating the complexity of tuning in V1 neurons.} 

\section*{Materials and Methods}

\subsection*{Ethics statement}

\revise{All procedures involving animals for generating the data of this paper were in accordance with the Guide of Institutional Animal Care and Use Committee (IACUC) of Peking University Animals, and approved by the Peking University Animal Care and Use Committee (LSC-TangSM-5). }

\subsection*{Dataset: stimuli and neuronal responses}

The neuronal data studied in this paper were published in a previous study~\cite{tang}. They are the responses of neurons in the V1 superficial layer of two awake (Macaca mulatta) monkeys A and E, with respect to a set of stimuli. They are obtained using large scale two-photon imaging with the calcium indicator GCaMP5s as reported in~\cite{tang}. The calcium signals in response to visual stimuli of 1142 neurons from monkey A and 979 neurons from monkey E are imaged during a fixation task. The response of a cell is computed as the standard $\Delta F/F_0$, based on the averaged activity within an ROI (region of interest) during stimulus presentation in each trial. 

\begin{figure}[htp]
  \centering
  \includegraphics[width=8.52cm, height=6.0cm]{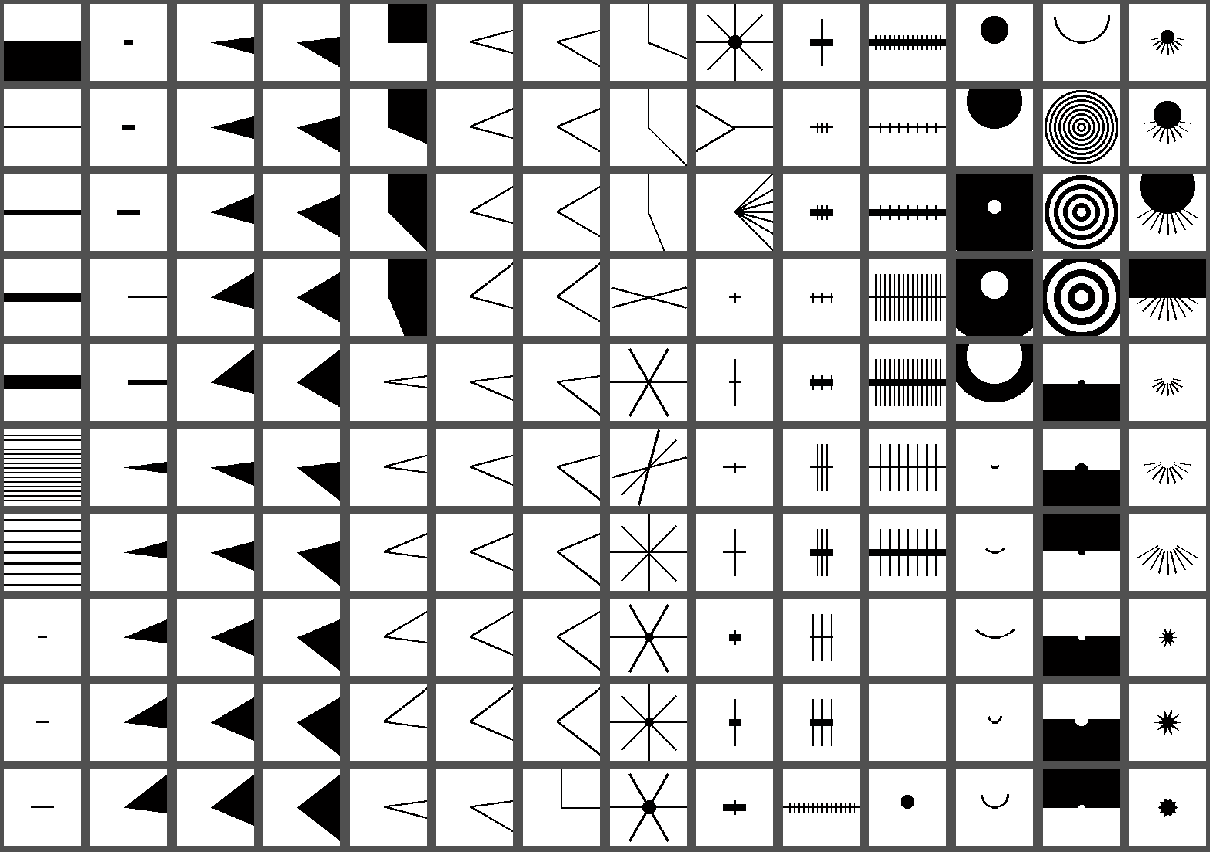}
  \caption{{\bf Image stimuli:} The 138 prototype stimuli, including simple orientation patterns as well as complex ones, which generate all 9,500 stimuli used through rotation and translation.}
  \label{fig2}
\end{figure}

The stimulus set is designed to test the hypothesis that many
neurons in V1 are not only orientation-tuned (OT) but also selective
to specific high-order (HO) complex features. The stimulus set contains 9500 binary (black and white) images generated from 138 basic prototypes, shown in Fig~\ref{fig2} by rotating and scaling. These prototypes are grouped into five major categories: 
orientation stimuli (bars and gratings), curvature
stimuli (curves, solid disks, and concentric rings),
corner stimuli (lines or solid corners), cross stimuli
(lines crossing one another), and composition
stimuli (patterns created by combining multiple elements
from the first four categories). 
The entire stimulus set consists of 1,600 OT patterns and 7,900 more complex HO patterns. The full set was shown to monkey A while a half-set (with half of the rotations) was shown to monkey E. Note that even for the half set, oriented bars at 48 orientations with rotation increment of $7.5^{o}$ were tested. 

Each stimulus is shown in a $3^{o} \times 3^{o}$  (90 x 90 pixels) aperture. The neurons possess overlapping receptive fields, all within a radius of $0.5^{o}$ in visual angle. All recorded neurons have classical RFs of diameters
well below $1^{o}$ in visual angle around the
stimulus center~\cite{tang}. Hence, to reduce the number of parameters in the models, we crop the central 40 $\times$ 40 pixels of all stimuli and downscale them to 20 $\times$ 20 pixels as model input. The input size is at least twice the size of any neuronal RFs and thus covers all of them. In this study, we discard neurons without a significant preference for stimulus patterns. Specifically, we discard neurons that have less than 50 stimuli with response $\Delta F/F_0$ greater than 0.2, where 0.1 is one standard deviation of the measurement. This measure reduces our samples to 781 neurons for monkey A and 632 neurons for monkey E. \reviseold{We used this selection criteria to ensure there is a sufficient number of stimuli with responses two standard deviations above noise level. We assumed that is necessary for the approaches to work well. Most of the neurons we eliminated based on this criteria have a low response to all pattern stimuli, rather than being very sharply tuned \revise{(see Supplementary Information)}.}

\subsection*{Gabor code and \reviseold{complex-shape} sparse code front-end }

The key contribution of this paper is the use of neural sparse codes derived from natural scenes as a front-end for receptive field models trained to characterize and predict neuronal responses. According to sparse coding theory~\cite{olshausenovercomplete}, the number of neurons used for natural scenes encoding can determine the complexity and diversity of the neural sparse codes. For example, for an input patch of 12 x 12 pixels, each \revise{gray-level} image is a vector in 144-dimensional space, with each pixel representing one dimension. When the number of neurons allocated to this patch is equal to 144, the representation is called a complete representation; otherwise, the representation will be either undercomplete or overcomplete. It has been shown that when the resource constraint is complete \revise{or} undercomplete, the basis functions learned are mostly Gabor filters or Gabor wavelets~\cite{18}. But in the overcomplete scenario, the basis learned tends to exhibit greater diversity and \reviseold{complexity~\cite{18, olshausenovercomplete, CSC, 27, 28, olshausen2013highly}}. 

\begin{figure}[!h]
\centering
  \includegraphics[width=5cm, height=5cm]{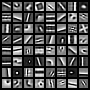}
  \includegraphics[width=3cm, height=5cm]{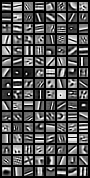}
  \includegraphics[width=5cm, height=5cm]{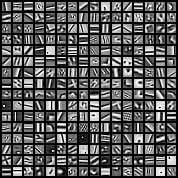}
\caption{{\bf \reviseold{Complex-shaped convolutional} sparse code dictionary:} Learned feature dictionaries from the CSC algorithm \revise{applied to grayscale natural images} with 64, 128, and 256 elements, each with a kernel size of 9px.
}
\label{fig1}
\end{figure}

\reviseold{In this study, we use the complex-shaped kernels obtained using the convolutional sparse coding (CSC) method~\cite{CSC}.} These are feedforward kernels that compute the coefficients of a set of complementary basis by linear projection for reconstructing the input images with the coefficients subject to sparsity constraint.
The use of convolution in sparse coding is shown to increase efficiency of patch-based sparse coding~\cite{CSC} that otherwise tends to produce Gabor filters or Gabor wavelets. CSC yields a variety of complex tunings reminiscent of the feature selectivity observed in~\cite{tang}. We train this unsupervised CSC model with a kernel size of 9 $\times$ 9 pixels on 1,000,000 25 $\times$ 25px image patches sampled from the 5000 128 $\times$ 128px natural scene images used in the original paper~\cite{CSC}. The model is trained with dictionaries of either 16, 32, 64, 128, or 256 channels. Some of the learned dictionaries are shown in Fig~\ref{fig1}. We observed that as we continue increasing the number of channels of the dictionary past a certain point, filters appear to have more redundancy instead of showing more diversified shapes. This model is complete in the sense that there are more than 625 neurons in the first hidden layer for an input image patch of 625 pixels. \revise{In early sparse coding works [16, 17], when convolution is not used, there is a greater amount of diversity of feature detectors in each hypercolumn. With convolution, overcompleteness can be achieved by a small set of feature detectors sampling the visual space more densely.  We adopt the convolution strategy to minimize the number of parameters in the model.}

%When convolution is not used, as in the earlier sparse coding works~\cite{18, 28}, there could be an even greater amount of diversity in the complex filters, but we use it anyway, since it intuitively corresponds better to the mechanisms of cortex.

\begin{figure}[htp]
  \centering
  \includegraphics[width=13cm]{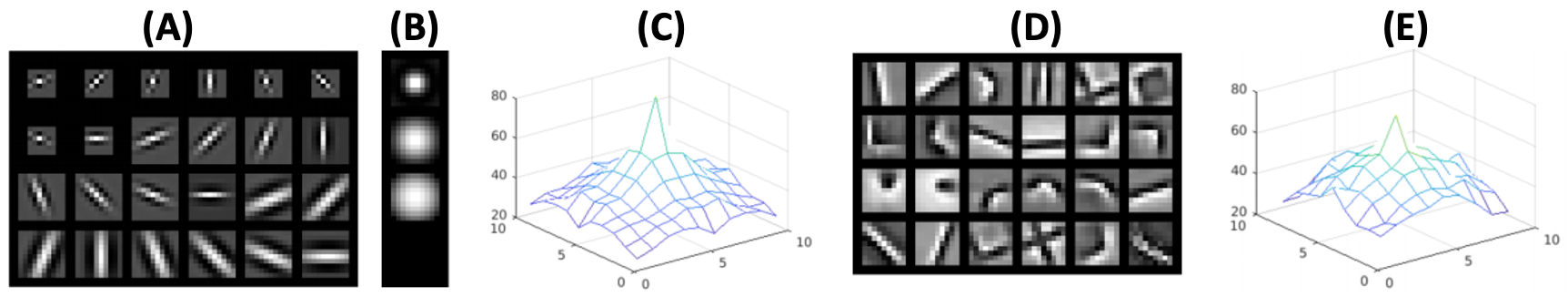}
  \caption{\reviseold{{\bf Gabor, Laplacian of Gaussian versus \reviseold{complex-shaped} sparse code front-ends:} From left to right we have: \textbf{A}: Gabor wavelets with 3 scales and 8 orientations. \textbf{B}. Laplacian of Gaussian kernels with 3 scales. \textbf{C}. The aggregated power spectrum of the Gabor wavelets. \textbf{D}: 24 \reviseold{complex-shaped} sparse kernels selected from the 64-element CSC model for use in our Fixed Kernel CNN and Projection Pursuit models. \textbf{E}: Aggregated power spectrum of the \reviseold{complex-shaped} sparse codes. }}
  \label{fig5}
\end{figure}

In our study, we compare the use of these \reviseold{complex-shaped} sparse code front-ends with Gabor wavelet front-ends \cite{daugman1988, lee96}, which have long been the mainstream models for V1 neurons. We also compare them against the Laplacian of Gaussian (LOG) filter front-end, which has been used to model retinal ganglion cells and V1 cells with center-surround RFs. Fig~\ref{fig5} shows an example of a set of Gabor~\cite{lee96} and LOG wavelets in contrast to a set of \reviseold{complex-shaped} sparse codes selected from the learned dictionary of the CSC algorithm. Note that the Gabor wavelet codes and the \reviseold{complex-shaped} sparse codes have similar coverage in the spatial frequency domain, as shown in the aggregated power spectrum in Fig~\ref{fig5}. \revise{Their spatial frequency spectra highly overlap with each other (see Supplementary Information)}, suggesting that their difference in performance is not likely due to different coverage in the frequency domain. 

\subsection*{Receptive field models for neuronal response prediction}

In this section, we study the effect of incorporating the \reviseold{complex-shaped} sparse code front-end into two classes of supervised learning models for modeling V1 neurons. The first class is the pursuit regression models, which perform layer-by-layer non-parametric regression. We extend the standard PPR approach~\cite{21, 23, 22} in two ways: first, by incorporating the convolution operation, and second, by incorporating the \reviseold{complex-shaped} sparse code as a front-end, i.e. using these codes as the feature kernels in the first layer of the model.  
The second class is CNNs. We compare baseline CNN models with kernels learned from scratch by a data-driven approach~\cite{zhang}, against models using fixed sparse code kernels in their first layer. One model of each type is learned for each neuron in monkey A and E. For fair comparison, we constrain the number of the parameters of the proposed models to be the same or lower than the baseline models. 

\smallskip
\noindent\underline{\textbf{Projection pursuit regression (PPR)}}~\cite{21} has been previously used to model neural responses, achieving remarkable performance~\cite{liu2016spatial} \reviseold{and allowing analysis of receptive field structure~\cite{20}}. This method does not require stimuli with specific statistical properties and is thus suitable for analyzing the neuronal responses to our complex pattern stimuli. The key idea of PPR is to optimize one kernel of the RF model at a time, allowing the first learned kernel to provide important information about the neuronal RF while the consecutive kernels only fit the residuals of the previous ones. This means that instead of the structure of preferred inputs being distributed evenly across many kernels, as in a standard CNN, it is concentrated as much as possible into a single, more \reviseold{structurally coherent} kernel.

The detailed PPR algorithm is shown in Algorithm~\ref{alg1}, where $\cdot$ denotes the dot product and \reviseold{$g(x)$ stands for a quadratic function $a*x^2 + b*x + c$ with learned parameters $a, b, c$}. This algorithm will sequentially learn layers of kernels $F$, which are the same size as the input image $x$. The regression of the neuronal response $y$ on the input stimulus $x$ is defined as the dot product between $x$ and the learned kernel, followed by a quadratic nonlinearity. After all layers have been learned, the kernels will be reordered based on pursuit contribution $I$, and redundant kernels will be eliminated. Finally, the function $\psi(S, F) = \sum_{m=1}^{|F|} g({F}_{m} \cdot S)$ will be used to predict the neuronal response with respect to input stimulus $S$.

\begin{algorithm}\
\small
\caption{PPR Algorithm}\label{alg1}
\begin{algorithmic}[1]
\Procedure{PPR}{$X,Y,M$}\newline \Comment{input set of images $X$, corresponding neuronal responses $y$, and number of kernels to be learned $M$.}
\State $\text{Define quadratic function } \reviseold{g(x) = a*x^2 + b*x + c}$
\State ${r}\gets {Y}, m\gets 1$
\While{$m\leq M$}
\State $I(F'_m) = 1 - \sum_{i=1}^{n} ({r}_{i} - g(F'_{m} \cdot X, 2))^2 / \sum_{i=1}^{n} {r}_{i}^2$
\State ${F}_{m}\gets argmax(I(F'_m))$
\State ${r}\gets {r} - \reviseold{g({F}_{m} \cdot X)}$
\State $m\gets m + 1$
\EndWhile\label{euclidendwhile}
\State $F \gets \{F_1, \cdots, F_M\}$ \\
\Return $\text{reduce\_redundant(F)}$
\EndProcedure
\end{algorithmic}
\end{algorithm}

\smallskip
\noindent Next, we introduce two variants of PPR that improve the neuronal prediction results and kernel interpretability. The first variant is
{\em Convolutional projection pursuit regression}, which introduces the convolution operation~\cite{lecun1995convolutional} to the kernel layers. The second variant is {\em Convolutional matching pursuit regression}~\cite{22}, which limits the feature kernels selected by the convolutional projection pursuit regression to a fixed dictionary of filters. 

\smallskip
\noindent\underline{\textbf{Convolutional projection pursuit regression (CPPR)}}:
The original PPR model requires the kernels $F$ to be the same size as the input images, leading to learning redundant parameters. More importantly, monkey V1 neurons' receptive field sizes are usually much smaller than image stimuli, and thus the kernels learned by PPR can not accurately describe the size of the neuronal RF. Here, we propose a modified version of PPR, CPPR, which allows using a smaller kernel of size $d\times d$ and performs convolution over the input images of size $w\times w$ at each layer. Then the result after convolution (of shape $(w-d+1) \times (w-d+1)$) goes through an \reviseold{absolute value layer} and a max-pooling layer to get the output, as shown in Equation~\ref{equ1}. All other procedures are the same as in Algorithm~\ref{alg1} except that we take convolution between kernels and the image instead of the simple dot product. Ideally, CPPR can learn kernels that are the same size as the receptive field, and the receptive field location can be chosen by the max-pooling layer. 

\begin{equation}\label{equ1}
    \psi_i(X, F) = \sum_{m=1}^{|F|} \reviseold{g \ ( \ max\ (\ abs({F}_{m} \ast X)))}
\end{equation}
In Equation~\ref{equ1}, $\ast$ denotes the convolution operation, $abs$ denotes the $absolute value$ layer and $max$ denotes the max pooling layer. 

\smallskip
\noindent\underline{\textbf{Convolutional matching pursuit regression (CMPR)}}:
We find that most of the kernels learned from PPR or CPPR algorithm tend to be rather noisy and lack coherent forms. \reviseold{We define the term ``noisy'' as having high power at high spatial frequencies.} Inspired by the matching pursuit regression algorithm~\cite{23}, we design CMPR methods to incorporate the \reviseold{complex-shaped} sparse code front-end into the CPPR models. Specifically, we modify the CPPR approach by selecting the feature kernels $F$ exclusively from an existing dictionary, instead of learning them from scratch. Here, we take the \reviseold{complex-shaped} sparse codes with 64 kernels (Fig~\ref{fig2}) and expand them eight-fold by rotating each component at $45^{o}$ step by step to allow better orientation steering of the kernels. The resulting dictionary $D$ has 512 kernels. If the dictionary $D$ spans the space which we intend to learn from CPPR, then CMPR should have similar performance to CPPR. More importantly, the kernels selected using the CMPR model will be much \reviseold{less noisy} than those learned by CPPR. We also tried to use \reviseold{complex-shaped} sparse codes without rotational expansions as the dictionary, but the performance was substantially worse.

\smallskip
\noindent\underline{\textbf{Convolutional neural networks (CNNs)}} are the state-of-the-art models for predicting neuronal responses~\cite{zhang, 24, 29, 31, Yamins2014, Yamins2016}. They outperform all other baseline models, such as Gabor-based
standard models for V1 cells and the various variants of generalized linear models~\cite{7, 8, 24, 29}.

\begin{figure}[htp]
  \centering
  \includegraphics[width=10cm]{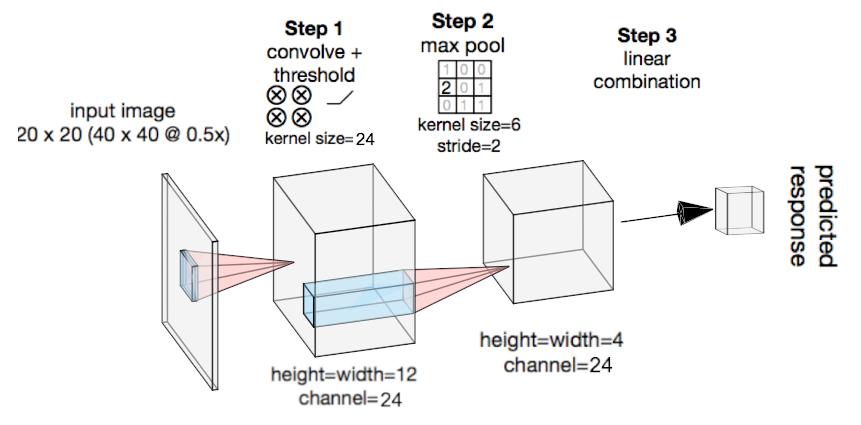}
  \caption{{\bf Structure of CNN:} 
  The input image is of size $20\times 20$ pixels, downsampled from the $40\times 40$ image of the experimental stimulus centered at the neurons' aggregated receptive fields. The model has three stages: convolution and thresholding, max-pooling, and a final fully-connected linear combination to generate the predicted neuronal response. \\
  }
  \label{fig4}
\end{figure}

We use the structure of an earlier paper~\cite{zhang}, which is optimized for this set of pattern data. A one-layer CNN model passes the input images through a series of linear-nonlinear operations, each of which consists of convolution, ReLU nonlinearity, and max pooling (Fig~\ref{fig4}). Finally, the outputs of the above operations are linearly combined to form the predicted neuronal response. Here we experiment with two specific CNN structures also tested in~\cite{zhang} for comparison: one with 4 kernels in the convolutional layer ($CNN_4$) and another with 9 kernels ($CNN_9$). The earlier study~\cite{zhang} explored different CNN structures and established that adding more layers to the model does not increase the fitting ability of the model for this dataset, \revise{so we do not experiment with deeper models.} $CNN_9$ has been shown to be the state-of-the-art model for this dataset, superior to a variant of the projection pursuit methods. 

\smallskip
\noindent\underline{\textbf{Fixed kernel convolutional neural networks (FKCNNs)}}:
Even though the standard CNN models have the best predicting accuracy, the recovered features (i.e. the 4 or 9 kernels in the first layer) in the hidden layer often do not have \reviseold{coherent} forms, making it difficult to understand the constituent components of the neuronal RF. Here, \reviseold{ we explore whether  the use of the sparse code front-end can improve its performance by reducing overfitting and whether the constituent components of a neuron's receptive field can be related to the statistical priors of natural scenes.} 

FKCNNs use the baseline CNN model's structure (Fig~\ref{fig4}), but replace the first layer with either the sparse code dictionaries learned from the sparse coding algorithm (Fig~\ref{fig1}) or Gabor wavelets, which are considered standard receptive field models of V1. During training, the substituted first layer is kept fixed. In order to make a fair comparison with the standard CNNs, we fix the number of parameters learned by the FKCNNs to be roughly the same. Specifically, we select 24 distinct complex or Gabor kernels of size $9 \times 9$ as the first convolutional layer's kernels. This results in the same number of parameters to learn as the $CNN_4$ and half the number of parameters as in $CNN_9$, since the kernels themselves are not learned. The 24 kernels are selected in the following manner: we start with a set of 64 kernels learned with CSC (Fig~\ref{fig2}) and discard the kernels that are too noisy or redundant. Then we randomly subsample 20 sets of 24 kernels as the fixed first layer from the remaining 35 kernels. We construct FKCNN models from these sets of 24 kernels as well as the entire set of 35 kernels and tested their performance on the pattern dataset. The results have an average performance in terms of Pearson correlation of 0.445 with a standard deviation of 0.0079, comparable to the normal CNN models, suggesting that a random subset with size 24 from these 35 kernels is complete enough to model this dataset.

In addition to the \reviseold{complex-shaped} sparse code dictionary, we also test the Gabor wavelet ( Gabor-CNN or GCNN) as shown in Fig~\ref{fig5}. Since some V1 neurons are known to have center-surround receptive fields, we also add three Laplacian of Gaussian feature channels of different scales to supplement the Gabor wavelet dictionary (we denote it as GLCNN). The purpose of introducing GCNN and GLCNN here is to assess whether the \reviseold{complex-shaped} sparse codes provide a better dictionary for modeling V1 neurons than the standard models.

\noindent\underline{\textbf{\reviseold{Comparison of model architectures of the two approaches:}}} 

\reviseold{The PPR-based methods and the CNN-based methods we explored for neural response prediction  have the following three notable differences in their designs: a) for training, PPR-based methods train one kernel at a time, in contrast to CNN, which train all kernels and parameters simultaneously; b) for CNNs, the nonlinearity is ReLU, while PPR methods use the absolute value or quadratic non-linearity; c) for CNNs, the max pooling retains some spatial information, and PPR methods pool everything.} 

 \reviseold{These are the defining and intrinsic features of the two approaches. PPR emphasizes finding the primary components and then fitting the successive residuals while CNNs seek to find potentially many different ways to compose the receptive field.  As for the choice of nonlinearity, a quadratic nonlinearity is used in the original PPR method, giving PPR-based models three tunable parameters and thus extra degrees of freedom over a ReLU or sigmoid function. This is an intrinsic advantage of the PPR-based methods. In CMPR, the filters are fixed, and only the prediction scale and bias need to be adjusted; using ReLU or sigmond will deprive them of learnable parameters. As for the pooling choice, the key idea of the projection pursuit is to find the best filter at the best location, which max pooling addresses.  Retaining some spatial information in CNN-based methods confers more flexibility and expressiveness, allowing the CNN to synthesize the receptive fields by combining information across both feature channels and space. This is an intrinsic advantage of the CNN approach. Thus, even though in theory, the two approaches could use different pooling and nonlinearity strategies, our design choices are typical and suit the relative advantages of their respective methods.  }

\section*{Results}
\begin{figure}[htp]
  \centering
  \includegraphics[width=8cm]{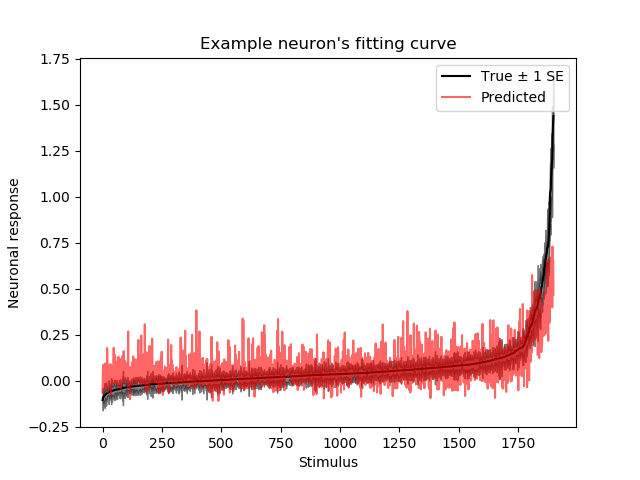}
  \caption{ \reviseold{{\bf Predicted tuning curve versus actual tuning curve:} An example of a real neuronal tuning curve to \revise{the 1900 stimuli from} the testing dataset (not used during training), ordered by the response magnitude (black curve, \revise{with $\pm$ 1 standard error estimated from the trial-by-trial variance}). The red curve shows the model's predicted response to these stimuli. The Pearson correlation (0.61 in this example) between these two curves is calculated as a quantitative evaluation metric.}}
  \label{fitting_curve}
\end{figure}

In this section, we first evaluate the performance of different models quantitatively and then discuss the model's interpretability from various aspects. \reviseold{In particular, our quantitative performance evaluation is based on neuronal response prediction accuracy, convergence rate, and data efficiency}. We use the Pearson correlation between predicted and actual neuronal responses as an evaluation metric to measure prediction accuracy, which is widely used in many papers~\cite{4, 24, zhang}. An example can be found in Fig~\ref{fitting_curve}.

\subsection*{Quantitative Evaluation of the Methods}

\begin{figure}[htp]
  \centering
  \includegraphics[width=9cm]{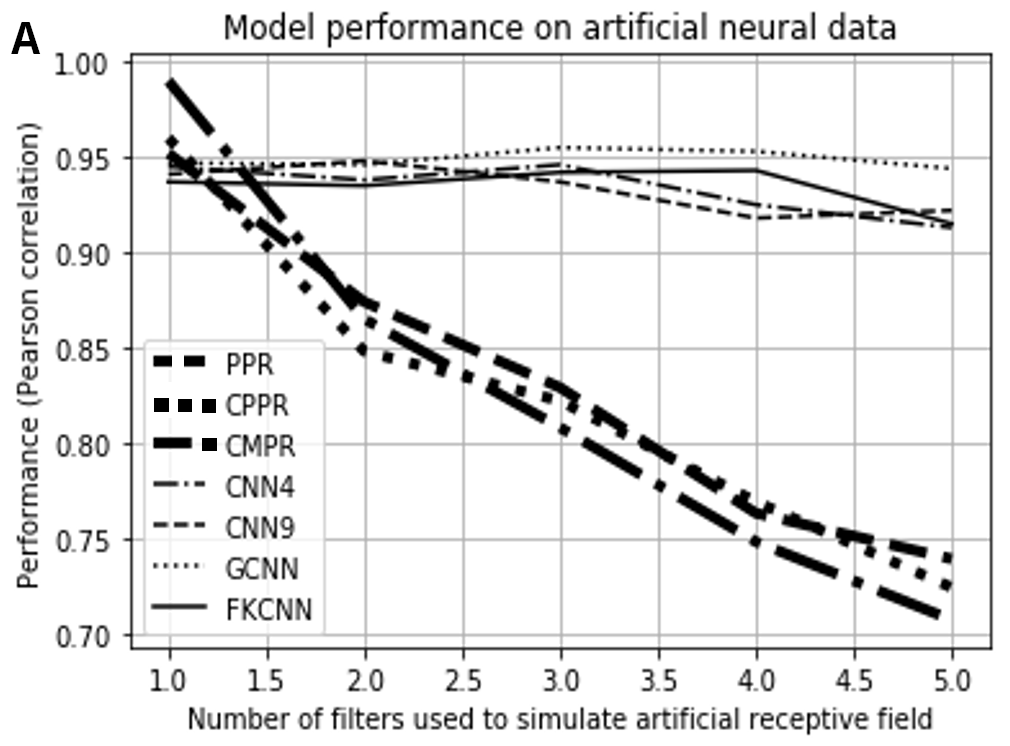}
  \includegraphics[width=11cm]{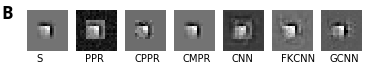}
  \includegraphics[width=5cm]{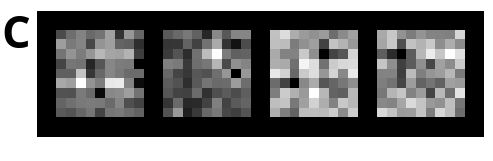}
  \includegraphics[width=5cm]{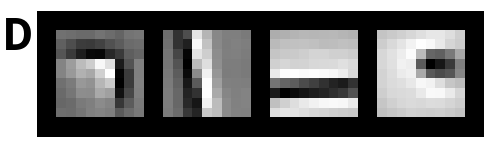}
  \caption{\reviseold{{\bf Model performance on artificial data:} This figure shows the performance (A) and the recovered receptive fields (B-D) of the different methods. On the second row (B) from left to right, the seven images are the receptive field $s$ recovered from each model type: PPR's filter, CPPR's filter, CMPR's selection, CNN's visualization, FKCNN's visualization, and GaborCNN's visualization. On the last row, the left images (C) are the learned kernels of the CNN. The right image (D) shows the top 4 highest-contribution kernels of the FKCNN model, \reviseold{whose importance scores are 0.782, 0.503, 0.469, and 0.391 respectively.}}}
  \label{fig6_2}
\end{figure}

\noindent\underline{\textbf{Synthetic Data:}} We evaluate these novel methods by first testing them using synthetic neurons. The synthetic neurons are modeled as the nonlinear transform of one or the sum of multiple linear filter outputs. We choose the ReLU activation function as our nonlinear transform because previous work found that all activation functions behave essentially the same~\cite{zhang}. Then, they are stimulated by a set of Gaussian white noise images (20 x 20 pixels), and the response is given by the following linear-nonlinear model, as described by Equation~\ref{equ2}. This is a common ground truth evaluation method, also used by~\cite{4, 24}.

\begin{equation}\label{equ2}
    y_i = \sum_{s \in S} ReLU(I_i \cdot s) + \eta
\end{equation}
where $\cdot$ denotes the inner dot product between the input image and the artificial RF $s$; $ReLU$ is the rectified linear unit function that maps negative values to 0; $\eta$ represents Gaussian noise with zero mean and variance equal to $\frac{\sqrt{I_i \cdot s}}{10}$; the set of artificial RFs $S$ is a subset of $ 9 \times 9$ filters padded with zeros in the surrounding to make them the same size as the input stimulus ($20 \times 20$). We select $S$ either from our learned \reviseold{complex-shaped} sparse codes or cropped patches of our pattern stimuli. 
We train the models with 10000 samples of noise inputs $I$ until convergence, and then test them with another set of 2000 noise input. As shown in Fig~\ref{fig6_2}, for these simple neurons, all the approaches described in the previous section are able to achieve near perfect testing performance when we only have one RF in $S$. However as we gradually increase the number of filters in $S$, the pursuit regression models' performance seems to suffer whereas the CNN models roughly retained their good performance. 
Fig~\ref{fig6_2} shows one example of the recovered receptive fields using different pursuit regression models and CNN's visualization~\cite{26}. Even though the visualization of the CNN model of the neuron is consistent with $S$, the recovered kernels in the hidden layer are very noisy and impossible to interpret, whereas the FKCNN's component kernels  \reviseold{are statistical priors of natural scenes with coherent structured patterns}. Specifically, panel D of Figure~\ref{fig6_2} demonstrates the top 4 ranked kernels of FKCNN method. \reviseold{We can see that the first kernel (the most important one) matches with the artificial RF. In fact, the importance value of the first kernel, is much higher than those of the rest kernels. Therefore in this case, we have reasons to believe that the first kernel resembles the RF.
For the second top-ranked kernel, it matches with the vertical bar in the RF. Therefore, it can partially response to testing stimulus and derive neuronal responses that has a relatively high correlation with the actual neuronal responses.}

\begin{figure}[htp]
  \centering
  \includegraphics[width=6.5cm]{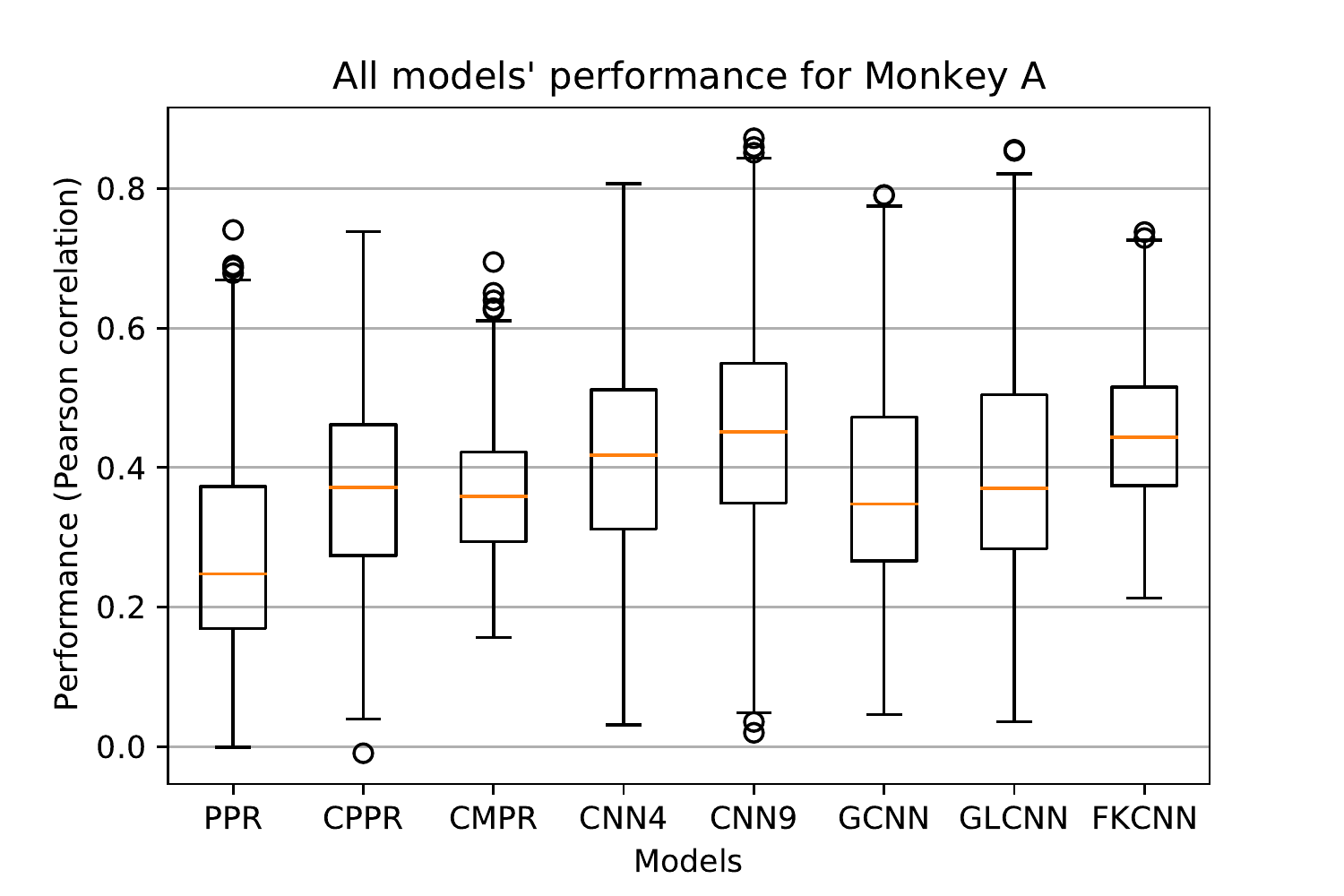}
  \includegraphics[width=6.5cm]{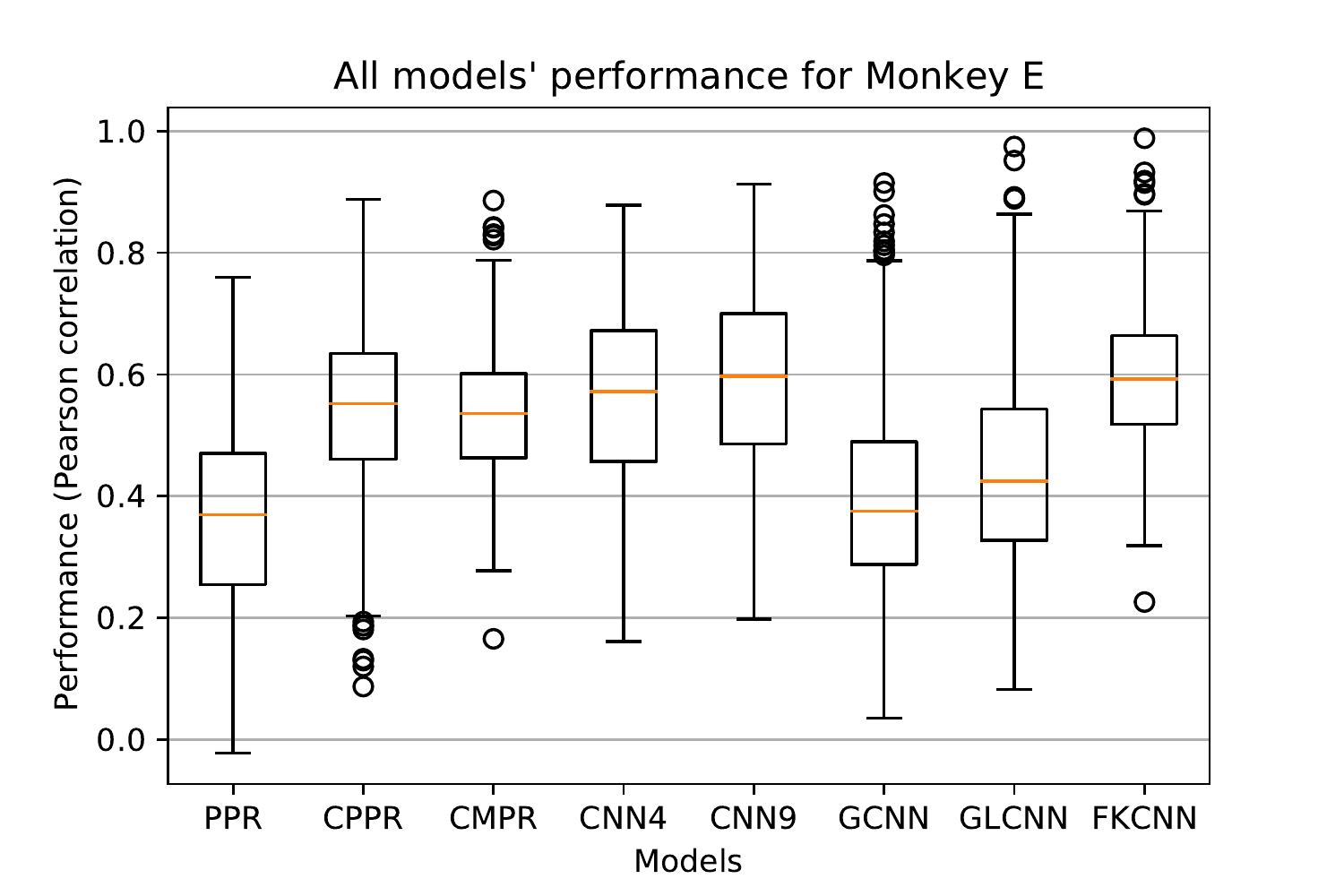}
  \caption{{\bf Model performance on neuronal data:} \revise{Box-plots of response prediction performance based the Pearson correlation  between true and predicted responses of all modeled neurons. The performance of all models on Monkey A is shown on the left and performance on Monkey B is shown on the right.} Overall, the performance is significantly better for Monkey E because of higher-quality neuronal recordings for this monkey.}
  \label{fig8}
\end{figure}

\smallskip
\noindent\underline{\textbf{Real Neuronal Data:}} We test all seven methods on the calcium imaging neuronal data.
For the three projection pursuit based approaches, all models can select up to 10 kernels. The CMPR model was tested using a 512 feature dictionary derived from 64 \reviseold{complex-shaped} sparse code filters learned from CSC (Fig~\ref{fig1}), with 8 rotations each. The CMPR and CPPR models have roughly the same number of parameters to learn, which is about 25\% of that used in PPR. It is worth noting that CMPR takes longer to train because of the selection process. Fig~\ref{fig8} shows that CPPR and CMPR achieve prediction accuracy better than PPR with fewer parameters, which underscores the efficacy and efficiency of the convolution approach~\cite{zhang}. 

\begin{table}[!ht]
    \centering
    \scalebox{0.78}{
    \begin{tabular}{c|c|c|c|c|c|c|c|c}
    \hline
        Model & PPR & CPPR & CMPR & $CNN_4$ & $CNN_9$ & GCNN & GLCNN& FKCNN  \\ \hline
        
        PPR & [0.278] &-32.7\%$^{***}$ & -30.9\%$^{***}$ &-48.9\%$^{***}$ & -61.2\%$^{***}$ & -35.6\%$^{***}$ & -42.1\%$^{***}$ & -62.9\%$^{***}$ \\ \hline
        
        CPPR & 32.7\%$^{***}$ & [0.369] & 1.3\% & -12.2\%$^{*}$ & -21.4\%$^{**}$ & -2.2\% & -7.0\%$^{*}$ & -22.8\%$^{**}$ \\ \hline
        
        CMPR & 30.9\%$^{***}$ & -1.3\% & [0.364] & -13.7\%$^{*}$ & -23.0\%$^{**}$ & -3.6\% & -8.5\%$^{*}$ & -24.5\%$^{**}$ \\ \hline
        
        $CNN_4$ & 48.9\%$^{***}$ & 12.2\%$^{*}$ & 13.7\%$^{*}$ & [0.414] & -8.2\%$^{*}$ & 9.8\%$^{*}$ & 4.8\%$^{*}$ & -9.4\%$^{*}$ \\ \hline

        $CNN_9$ & 61.1\%$^{***}$ & 21.4\%$^{**}$ & 23.0\%$^{**}$ & 8.2\%$^{*}$ & [0.448] & 18.8\%$^{**}$ & 13.4\%$^{**}$ & -1.1\% \\ \hline
        
        GCNN & 35.6\%$^{***}$ & 2.1\% & 3.5\% & -9.8\%$^{*}$ & -18.8\%$^{**}$ & [0.377] & -4.8\%$^{*}$ & -20.2\%$^{**}$ \\ \hline
        
        GLCNN & 42.1\%$^{***}$ & 7.0\%$^{*}$ & 8.5\%$^{*}$ & -4.8\%$^{*}$ & -13.4\%$^{*}$ & 4.8\%$^{*}$ & [0.395] & -14.7\%$^{**}$ \\ \hline
        
        FKCNN & 62.9\%$^{***}$ & 22.8\%$^{**}$ & 24.5\%$^{**}$ & 9.4\%$^{**}$ & 1.1\% & 20.2\%$^{**}$ & 14.7\%$^{***}$ & [0.453]  \\ \hline

    \end{tabular}
    }
    \caption{
    \revise{
{\bf Relative performance difference between the different methods on Monkey A:} The average performance of all models are put on the diagonal of this table.
The relative performance difference is calculated as (A-B)/min(A, B), where A is the average performance of the method on the row and B is that of the method on the column.\\
* refers to a $p$-value between 1e-10 and 0.05 \\
** refers to a $p$-value between 1e-40 and 1e-10 \\
*** refers to a $p$-value between 0 and 1e-40.\\
Otherwise, it means that the difference between two methods is not significant.}
}
    \label{tab1}
\end{table}

Fig~\ref{fig8} and Table~\ref{tab1} show that our FKCNN model performs significantly better than the $CNN_4$ model which has the same number of parameters and comparable to $CNN_9$ model (state-of-the-art for this set of data~\cite{zhang}) for both monkeys. The fact that the FKCNN's \reviseold{complex-shaped} sparse code dictionary serves as an effective front-end suggests that the \reviseold{complex-shaped} sparse code might indeed span a similar or even better feature space than that spanned by the noisy constituent components of CNN for characterizing the real neurons.

\begin{figure}[htp]
  \centering
  \includegraphics[width=10cm]{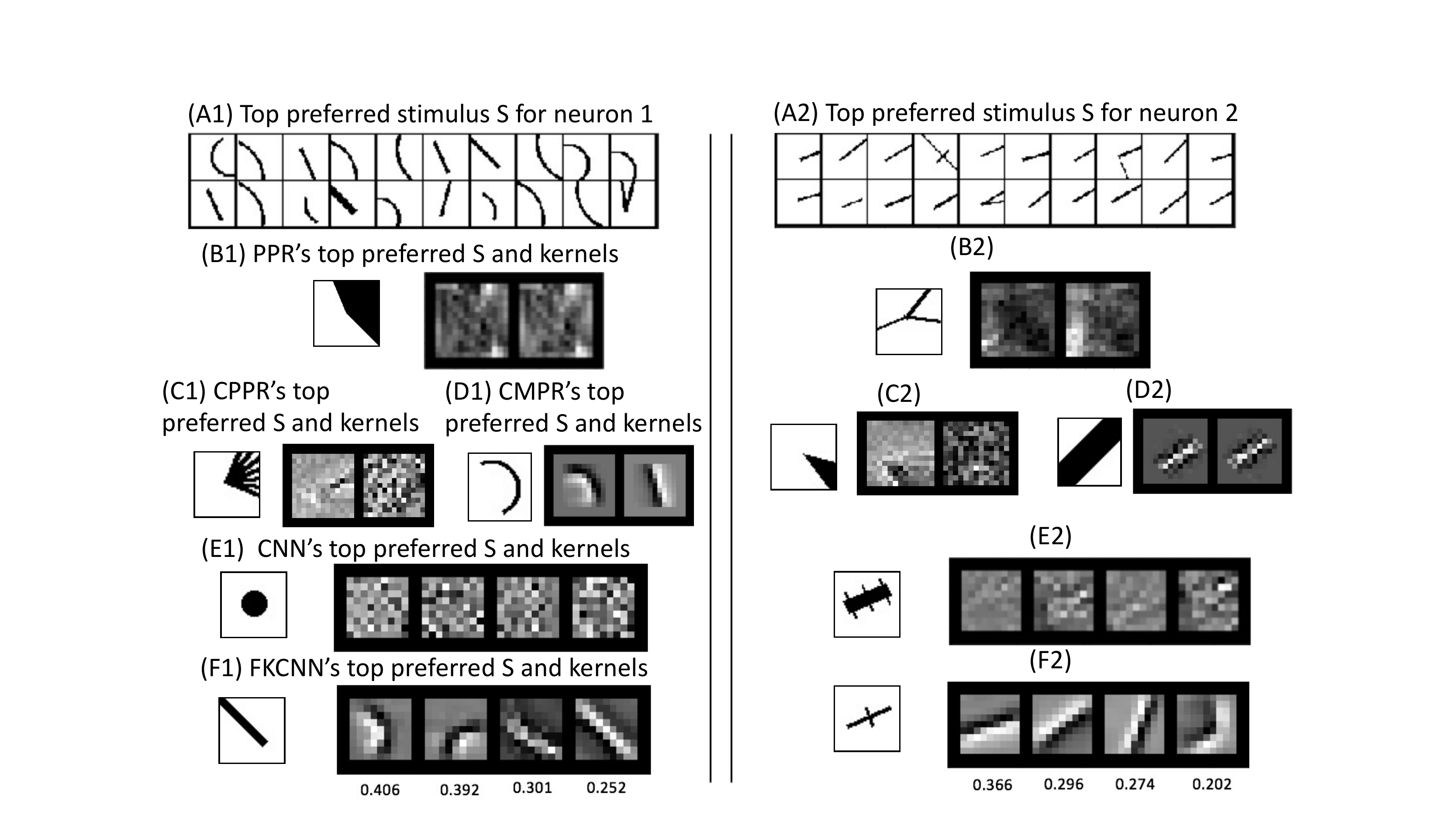}
  \caption{{\bf The recovery of two example neurons' receptive field components:} Two neurons are shown in the figure, separated by the double lines. Images (A1) and (A2) show each neuron's top 20 preferred stimuli (size 40x40). \revise{The most preferred stimulus and the two kernels (size 20x20) of PPR are shown in images (B1) and (B2) for two example neurons respectively. The most preferred stimulus and the two kernels of CPPR are shown in images (C1) and (C2), and those of CMPR are shown in images (D1) and (D2). The most preferred stimulus and four learned convolutional kernels from $CNN_4$ (size 9x9) for each neuron are shown in images (E1) and (E2). The most preferred stimulus and four highest-importance kernels in the FKCNN model (size 9x9) and their importance scores underneath the kernels are shown in images (F1) and (F2).}}
  \label{fig9}
\end{figure}

\smallskip
\noindent\underline{\textbf{Interpretation of Recovered Receptive Field Components:}}
The CMPR and CPPR models have comparable performance, but the receptive fields recovered by the CPPR method often do not have an \reviseold{coherent structure}. \reviseold{Fig~\ref{fig9} compares the kernels recovered by the standard PPR method (Images (B1) and (B2)), those recovered by CPPR (Images (C1) and (C2)), and those recovered by CMPR (Images (D1) and (D2)) for two example neurons. The PPR and CPPR methods yield filters that are noisy, while CMPR reveals the understandable key component features preferred by the neurons.}

\reviseold{In Fig~\ref{fig9}, the kernels from the CNN model's convolution layer (Images (E1) and (E2)) appear to be noisy and lack coherent structure, whereas the top FKCNN's component kernels are %more \reviseold{coherent} as 
curves and edges(Images (F1) and (F2))}. We rank the FKCNN's 24 kernels by their importance scores and showed the top four kernels in Fig~\ref{fig9}(fourth row). \revise{The importance score of each kernel component in Fig~\ref{fig9} is the response prediction performance on the testing data performance in terms of Pearson correlation when we retain only that filter and delete all other filters in the FKCNN after the training process}. 
\revise{We can see that the top-ranked FKCNN kernels (shown in F1 and F2 of Fig~\ref{fig9}) are reasonable components for constructing the top preferred stimuli (shown in A1 and A2 of Fig~\ref{fig9}).} \revise{ While the receptive fields of the neurons are difficult to visualize because of the non-linearity,  the most preferred stimulus extracted from the entire stimulus set by each model is a reasonable reflection of the receptive fields characterized by the model. We show the most preferred stimulus of each model to the left of the top kernels of each of the models. We observe that the top stimulus extracted by CMPR and FKCNN are indeed among the top 3 preferred stimuli of these two neurons, while models based on PPR, CPPR and CNN seem to have identified less relevant features.}  

%The \reviseold{test set correlations} of the FKCNN are 0.668 for neuron on the left and 0.587 for neuron on the right.

%(Figure 7: All models' performance on pattern data.)

\subsection*{\reviseold{Data efficiency and convergence rate}}

\begin{figure}[htp]
  \centering
  \includegraphics[width=4cm]{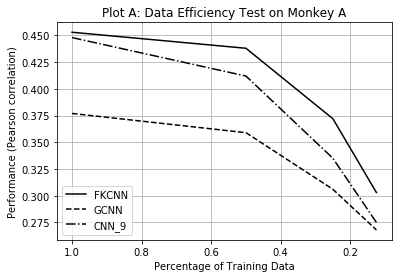}
  \includegraphics[width=4cm]{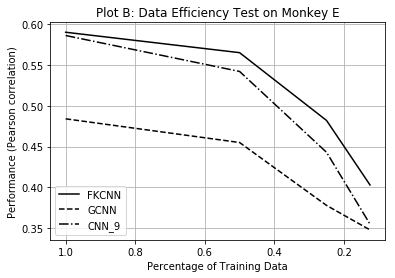}
  \includegraphics[width=4cm]{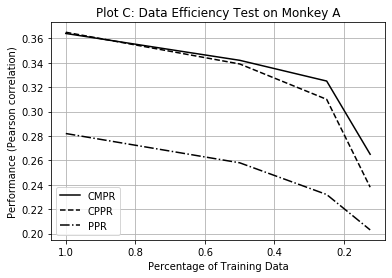}
  \includegraphics[width=4cm]{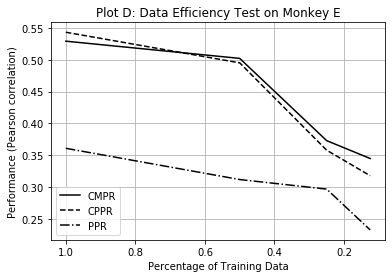}
  \includegraphics[width=4cm]{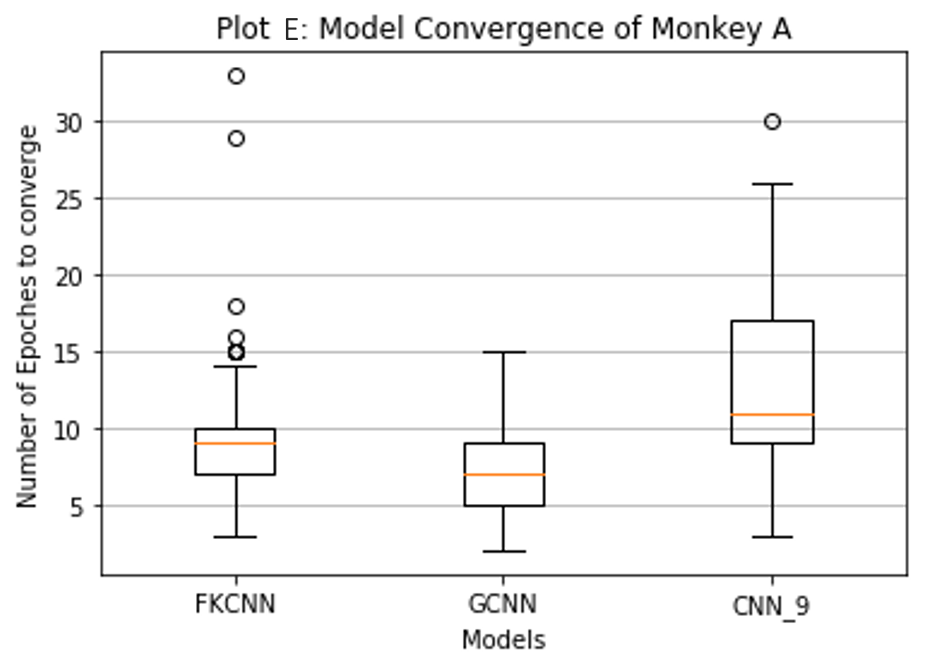}
  \includegraphics[width=4cm]{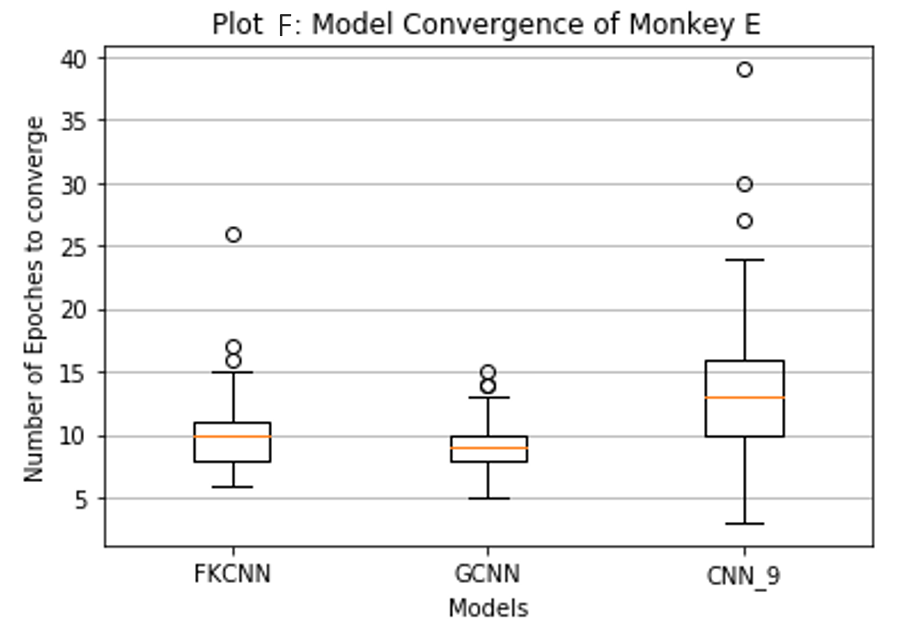}
  \caption{\reviseold{{\bf Model data-efficiency and convergence test:} Plots A, B, C and D show model performance as a function of the proportion of data used for each monkey and model type. Plots E and F summarize the distributions of training times (measured in epochs) across neurons for each model type and monkey in box-and-whisker plots. Please note that we did not compare the PPR, CPPR, and CMPR's convergence rate because they each have different training procedures, which don't allow fair comparison.}}
  \label{fig11}
\end{figure}

\reviseold{In this section, we compare the FKCNN, GCNN, CPPR, and CMPR with the baseline CNN and PPR models on two additional aspects beyond the already-discussed test performance: data efficiency and convergence rate. The results are shown in Fig~\ref{fig11} and the statistical significance (tested with paired t-test) of performance differences is reported in Table~\ref{tab-fig9}}.

\smallskip
\noindent\underline{\textbf{Data efficiency:}} \reviseold{In order to test the data efficiency of different models, we randomly subsample a fraction (100$\%$, 50$\%$, 25$\%$, or 12.5$\%$) of the training set to train the model while keeping the validation and test sets fixed to evaluate the model performance. Then we measure the drop of model performance with regard to the performance trained with $100\%$ and report this drop and statistical significance of Monkey A only in Table~\ref{tab-fig9}, because we observe the same pattern for Monkey E.}
\reviseold{From Plot A and B, we can see that FKCNN and GCNN require less data than the baseline $CNN_9$ to achieve the same or better predictive performance.} 
\reviseold{From Plot C and D, we can see that CMPR requires less data than CPPR to achieve the same or better predictive performance.} 

\smallskip
\noindent\underline{\textbf{Convergence rate:}} FKCNN and GCNN converge much faster than $CNN_9$ during training, i.e. the number of epochs needed for the model to have stabilized test performance is fewer than that of $CNN_9$. The models with the fixed front-end converge faster and require less data during training because the gradients do not need to propagate to the first convolutional layer as opposed to the baseline CNN models.

\begin{table}[h]
    \centering
    \scalebox{0.9}{
    \reviseold{
   \begin{tabular}{|l|l|l|l|l|l|l|l|l|}
\hline
{\bf Model} & \multicolumn{3}{|l|}{\bf Performance Drop} & \multicolumn{3}{|l|}{\bf T-test of averaged difference}\\ \hline
 & $50\%$ Data & $25\%$ Data & $12.5\%$ Data & FKCNN & GCNN & CNN$_9$ \\ \hline
FKCNN & $3.3 \%$ & $17.9 \%$ & $33.1 \%$ & / & - & +++ \\ \hline
GCNN & $4.7 \%$ & $18.8 \%$ & $28.9 \%$ & + & / & +++ \\ \hline
CNN$_9$ & $8.1 \%$ & $25.2 \%$ & $38.6 \%$ & - - - & - - - & / \\ \hline
\end{tabular}}
}

\scalebox{0.9}{
\reviseold{
\begin{tabular}{|l|l|l|l|l|l|l|l|l|}
\hline
{\bf Model} & \multicolumn{3}{|l|}{\bf Performance Drop} & \multicolumn{3}{|l|}{\bf T-test of averaged difference}\\ \hline
 & $50\%$ Data & $25\%$ Data & $12.5\%$ Data & PPR & CPPR & CMPR \\ \hline
PPR \  & $13.6 \%$ & $17.6 \%$ & $35.5 \%$ & / & +++ & - \\ \hline
CPPR \  & $8.8 \%$ & $34.0 \%$ & $41.4 \%$ & - - - & / & - - - \\ \hline
CMPR \  & $5.1 \%$ & $27.4 \%$ & $35.8 \%$ & + & +++ & / \\ \hline
\end{tabular}}}
    \caption{\reviseold{\textbf{The performance drop when using less data on Monkey A.} For statistical significance, we adopt the same notation from Table~\ref{tab1}.}}
    \label{tab-fig9}
\end{table}

\subsection*{Assessment of Higher Order Selectivity}

\noindent\underline{\textbf{HO versus OT selectivity:}} The standard models of V1 receptive fields are Gabor filters~\cite{1, 2}. Yet, we find that the FKCNN with a \reviseold{complex-shaped} sparse code front-end outperforms that with a Gabor front-end, even with Laplacian of Gaussian filters added (Fig~\ref{fig8}). In an earlier paper~\cite{tang}, we classified these neurons into simple orientation tuned (OT) and \reviseold{higher-order tuned} (HO) groups using a very stringent criterion. Specifically, a neuron is categorized as OT if there exists one or more single oriented bar or edge stimuli among the set of stimuli that elicit a response above 50\% of the peak response of that neuron to the entire stimuli set. Thus, the neuron can be considered to exhibit higher-order (HO) selectivity only when all the preferred stimuli are HO. About 40\% of the neurons in the superficial layer of V1 are classified as HO in this manner. The most preferred patterns of these neurons highly resemble the dictionary filters learned from the sparse coding algorithm \reviseold{(some examples of this can be found in section 3 of the Supplementary Information)}. Thus it is intuitive to conjecture that the \reviseold{complex-shaped} sparse code FKCNN would outperform the Gabor-based one (GCNN) by a larger margin for cells classified as HO than those classified as OT. Table 3 shows that to be the case on average, with the relative gain from the complex FKCNN over the GCNN being 19\% higher for HO cells in Monkey A, and 18\% higher for Monkey E. This correspondence provides supporting evidence for the previous work's claim that HO neurons' receptive fields are composed of more complex feature elements.

\begin{table}[h]
    \centering
    \scalebox{0.9}{
    \reviseold{
   \begin{tabular}{|l|l|l|l|l|l|l|l|l|}
\hline
\multicolumn{1}{|l|}{\bf } & \multicolumn{2}{|l|}{\bf Monkey A} & \multicolumn{2}{|l|}{\bf Monkey E}\\ \hline
$Model$ & FKCNN & GCNN & FKCNN & GCNN\\ \hline
OT+HO neurons & $0.418 \pm 0.0037 $& $0.362 \pm 0.0036$ & $0.550 \pm 0.0043$ & $0.483 \pm 0.0041$\\ \hline
OT neurons & $0.424 \pm 0.0038$ & $0.389 \pm 0.0032$ & $0.588 \pm 0.0045$ & $0.544 \pm 0.0037$\\ \hline
HO neurons & $0.408 \pm 0.0035$ & $0.319 \pm 0.0040$ & $0.488 \pm 0.0041$ & $0.387 \pm 0.0047$ \\ \hline
\end{tabular}}}
    \caption{\reviseold{{\bf Average and standard error model performance on neurons with different types.}}}
    \label{tab2}
\end{table}

To further probe the greater improvement for the complex front-end for HO neurons, beyond the average difference, we propose the following complexity score to test whether a neuron's higher-order selectivity can be predicted by the level of that improvement. This score measures the superiority of the complex basis over the Gabor-based basis for a given neuron.

\begin{equation}\label{equ3}
    score(neuron) = \frac{Corr_{FKCNN}(neuron) - Corr_{GCNN}(neuron)}{Corr_{FKCNN}(neuron) + Corr_{GCNN}(neuron)}
\end{equation}

\begin{figure}[htp]
  \centering
  \includegraphics[width=6cm]{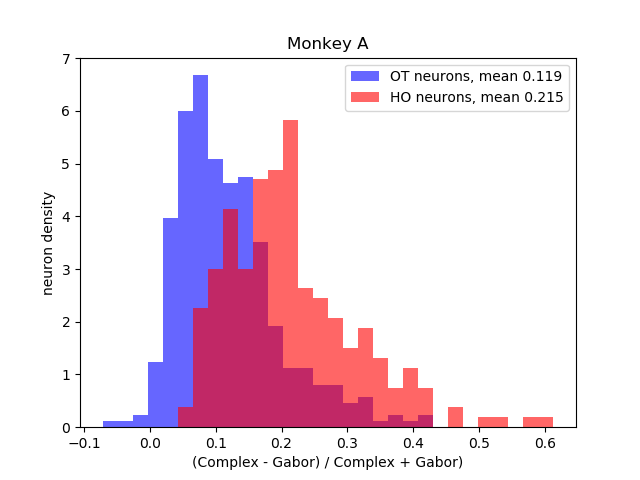}
  \includegraphics[width=6cm]{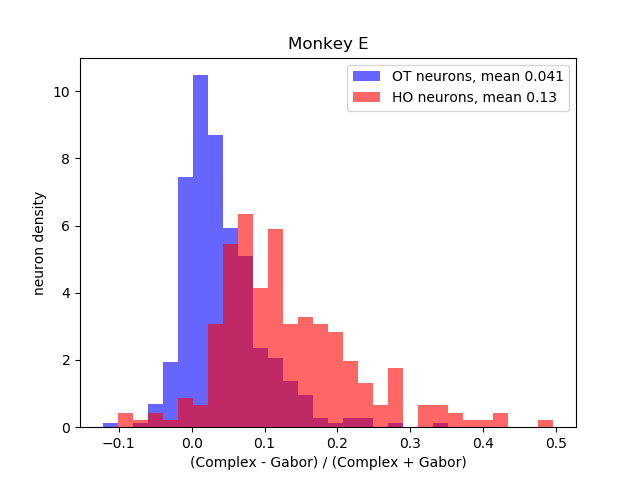}
  \caption{{\bf Distributions of HO and OT neurons' scores:} These figures show histograms of the HO and OT neurons' scores, measured by the difference in performance between the sparse FKCNN and Gabor models, for each monkey. Note that since the classes have different numbers of neurons, the histograms have been normalized by the total number of neurons in each group. The two cell types are fairly separable by this metric\reviseold{, with a linear separability of 73\% for Monkey A and 75\% for Monkey E, compared to baselines of 62\% and 61\% respectively.}}
  \label{fig12}
\end{figure}

We find that the HO and OT neurons indeed form visibly distinct distributions (Fig~\ref{fig12}) under this measure (equ~\ref{equ3}). A linear classifier trained to distinguish them based only on the measure achieves higher-than-chance, but not perfect accuracy (73\% for Monkey A, 75\% for Monkey E, compared to baselines of 62\% and 61\% respectively). These results demonstrate there is consistency between the complexity score with our earlier classification of HO and OT groups ~\cite{tang}. It is unlikely that either method perfectly corresponds to the ground truth of neuronal tuning, but their consistency despite very different methodologies suggests that some such complex tuning does exist. The FKCNN-based metric introduced here has the additional benefit of efficiency: it does not require extensive testing with a specific set of thousands of complex pattern stimuli, rather just enough data to fit the two types of model.

\begin{figure}[htp]
  \centering
  \includegraphics[width=13cm]{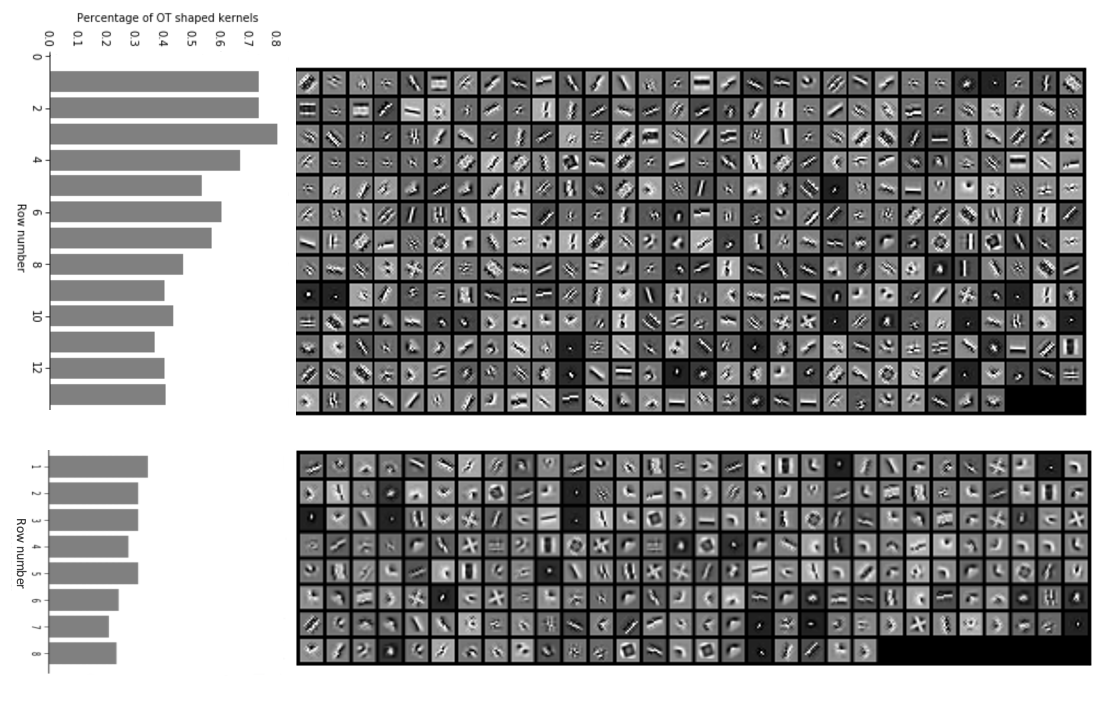}
  \caption{{\bf Classification of OT and HO neurons:} The figure shows the RF recovered by the CMPR algorithm for all 387 OT neurons and 232 HO neurons. The order is ranked by score, as described in equation~\ref{equ3}, in increasing order. Along the left margin, the proportion of each row with simple orientation kernels selected by CMPR is shown, clearly decreasing as the score increases, showing a correspondence between these metrics of HO tuning.}
  \label{fig10}
\end{figure}

\reviseold{We further explored the consistency of the original HO/OT classification with the top kernel extracted by the CMPR method.} 
We split the OT and HO neurons up, rank each group by its FKCNN/GCNN score in increasing order, and show each neuron's primary CMPR kernel  (Fig~\ref{fig10}). The CMPR kernels can be classified into simple oriented and complex groups, and the proportion of oriented filters in each row is shown on the left side, clearly decreasing as the FKCNN models increasingly outperform the GCNN models, and lower on average for the HO neurons. This metric, an estimate of the probability of CMPR selecting a complex filter rather than a simple oriented one, is again consistent with the other two, and is qualitatively matched by the appearance of the filters themselves as well. The findings in \cite{tang} thus are doubly reinforced by our basis-selecting methods, both in the comparison between a \reviseold{complex-shape basis}  and a Gabor-based one for CNN modeling, and the specific nature of the filters chosen from the complex basis for CMPR modeling.

\section*{\reviseold{Discussion}}

\reviseold{In this paper, we explore the novel idea of incorporating a sparse code dictionary as a front-end to the projection pursuit regression and convolutional neural network approaches for neural response prediction and receptive field characterization of neurons in the primary visual cortex of macaque monkeys. 
We show that the diverse \reviseold{complex-shaped} sparse codes learned from natural scenes based on the convolutional sparse coding principle can serve as an effective front-end to improve the predictive performance of both pursuit regression and CNN models.
Imposing these dictionary front-ends on RF models not only improves predictive performance, but also leads to faster convergence and higher data efficiency.} \reviseold{More importantly, it allows us to interpret the constituent components of a neuron's receptive field in terms of the dictionary or statistical priors learned from natural scenes using sparse coding theory.} 

\reviseold{It is important to note that the set of dictionary elements that can yield good performance is not unique. Different subsets of kernels drawn from the learned sparse codes yield comparable results as long as there is sufficient diversity in the kernel elements. Indeed, adding three center-surround (Laplacian of Gaussian) kernels to the mix of Gabor-wavelet kernels is found to yield a significant improvement in performance. We also found subsets of diverse kernels in the first-layer of the classical CNN AlexNet \cite{alexnet}  trained for image classification performs worse than  the complex-shaped sparse codes, but better than standard Gabor wavelets (see Supplementary Information). This suggests that convolutional methods learned with either a sparse coding objective or an image classification objective yield a diverse set of kernels that  span well the manifold of natural scene priors used by V1.
Given that the set of fixed kernels that can yield good performance is not unique, the kernel features \reviseold{selected} should be more appropriately considered as the basis for spanning the space of subcomponents of the neurons' receptive fields, rather than the actual subcomponents themselves. Nevertheless, these subcomponents give us a glimpse of the potential complexity and diversity of the cells' subcomponents.}

Compared to standard CNNs for neural response predictions~\cite{zhang, 29}, \reviseold{our approach provides information on the relative importance of more structured, coherent intermediate-layer kernels for predicting a neuron's response. This allows us to relate the constituent components of the neurons' receptive fields to the sparse codes learned from natural scenes, aiding our understanding over the noisy components learned by a standard CNN.}
 \reviseold{
 Efficient coding theory  \cite{18, olshausenovercomplete, 28, CSC, gregor2010learning}  predicts the emergence of receptive fields with coherent  features. 
It is an open question whether these coherently structured sparse codes provide intrinsic advantages for neural computation in the downstream visual areas.  
Here, we show that models with these sparse code kernels do generalize better, as evident by their data efficiency and faster convergence in learning.}   \reviseold{Perhaps,} having \reviseold{sparse code} kernels that  are meaningful visual features might also facilitate learning and inference in the nervous system \reviseold{as well}.

\reviseold{ The pattern stimuli we used in this study bear strong similarity to the complex-shaped sparse codes used in our front-end.  This stimulus set was designed to evaluate the hypothesis that  higher-order features known to be encoded in V2 and beyond are also encoded by V1 neurons~\cite{tang}. While it provides a great variety of higher-order features in different orientations and positions, it also features abundant orientation stimuli, over 1600 oriented edges, lines and gratings, in different lengths, widths,  end-point positions, and many orientations in step as fine as 7.5 degrees, projected to different positions within the 1-degree diameter visual receptive fields of the neurons located within a 500 um $\times$ 500 um cortical patch. Thus, we believe there is sufficient representation of orientation stimuli for our regression approach to select the oriented kernels, if they are appropriate.  The observation that the top kernels selected by the CMPR method for OT neurons are more likely to be oriented kernels than those selected for HO neurons (Figure~\ref{fig10}) reassures us of this assumption.} 

\reviseold{The high-order features such as corners, curvatures, junctions and crosses are given "equal opportunity" in our study \revise{ in that they are}  represented at a much higher rate than they are normally be observed in random natural scene patches. These higher-order features however were considered to be highly informative for vision \cite{attneave, Biederman91} and are widely assumed to be encoded by neurons in the extrastriate cortex~\cite{connorVanEssen, pasupathy2006}.  What is the consequence of this bias in the pattern stimuli tested on our findings regarding receptive field structure?  In the Supplementary Information, we provide results of these methods on \revise{the neurons of one monkey tested on} natural image stimuli. \revise{Natural images}  tend to have strong lower frequency components and feature predominantly uniform luminance regions or oriented edges in their random local patches.
Both the metrics based on FKCNN and the CMPR methods (Fig~\ref{fig12} and Fig~\ref{fig10} respectively), when trained on responses to natural images, fail to discriminate between the HO and OT neurons. These findings can be interpreted in two ways. One interpretation is that the pattern stimuli are important for detecting and discriminating the HO and OT neurons. \revise{ Natural images, when presented in 4X receptive field aperture as in our case, tended to evoke very sparse responses. When the amount of natural images is limited as in this case, the regression-based CMPR and CNN-based methods are not strong enough to detect the true nature of higher-order tuning in natural image responses. The second interpretation is that the particular higher-order tuning exhibited on the pattern stimuli, the preference for specific complex shapes like corners and crosses, is biased by the set of stimuli chosen, and may be different from the ground-truth higher-order tuning that produces the responses to natural images.
}}

\reviseold{Given these observations, our findings here alone do not constitute evidence for the diversity and complexity of V1 neurons. The arguments and evidence in support of selectivity and tuning of V1 neurons to higher-order features were put forth in \cite{tang}. Thus, our contribution here is primarily methodological. If we accept the premise that V1 HO neurons do exist, our approaches not only demonstrate some improvement in the CNN-based and PPR-based approaches in neural response modeling, but also provide useful metrics to detect  the HO neurons identified in the original study. We bolster that HO vs OT classification with two additional quantitative metrics, based on model properties instead of solely the neural response, and made possible by the use of fixed-kernel dictionaries in our models.  Recent iterative optimization approaches, such as the most exciting image (MEI) method~\cite{walker2019inception}, have also revealed complexity and diversity in V1 receptive fields in mice from natural images, even though earlier studies had suggested they have Gabor-like receptive fields.} \revise{This trend towards recognition of receptive field complexity is also present in work on the auditory cortex~\cite{auditory}.}

\reviseold{Our study shows that \reviseold{the diverse set of}   complex-shaped kernels provided by convolution sparse coding of natural scenes form an effective front-end for PPR-based and CNN-based approaches for characterizing V1 neurons' receptive fields in a data efficient manner. 
\reviseold{This finding shows that an overcomplete sparse code dictionary learned from natural scenes can potentially be related to the constituent elements of V1 receptive fields. This is not to say that they are the same, just that the sparse codes span the space of those elements better than the other alternatives such as Gabor filters, Alexnet filters or data-driven filters learned from scratch.} 
By making pairwise comparisons of the different methods, we gain some insight into the incremental benefit of each specific feature in the different methods.  The observation that CPPR is better than PPR indicates that max-pooling over convolution might provide an additional degree of freedom of matching flexibility, reminiscent of the classical deformable part models for object recognition~\cite{girshick}; that the FKCNN is better than CNN (and CMPR better than CPPR) suggests the features encoding natural scene priors provide a better embedding space; that the FKCNN is better than CMPR suggests that  compositional learning of multiple filters simultaneously in a neural network approach might further enhance matching flexibility to create a better embedding space; that the FKCNN is better than the GCNN suggests that a more diverse and complex set of kernels is important for modeling the responses of these V1 neurons, particularly those with higher order feature selectivity. \reviseold{Finally, the fact \revise{that} convolutional mechanisms are useful for learning the sparse codes and for providing flexibility in matching during model fitting in both the PPR-based and CNN-based methods suggests the potential relevance of convolution mechanisms for cortical information processing.}}

\section*{Acknowledgements}

This work was supported by an NSF grant CISE RI  1816568, and its Undergraduate Research Experience (URE) supplement, and the NIH grant R01 EY030226-01A1 to Lee, and National Natural Science Foundation of China (Grant No.31730109 and Grant No.U1909205) to Tang.
Ziniu Wu and Harold Rockwell were also supported by an interdisciplinary training fellowship in computational neuroscience at Carnegie Mellon sponsored by NIH NIDA grant 5T90 DA023426.

\nolinenumbers

\bibliography{refs}

%\section*{Supporting Information Legend for PCOMPBIOL-D-20-02146R2}

%Files S1. Supplementary Materials (plos\_revision\_supplementary.pdf)

%\noindent Text and figures describing additional experiments that support the claims in the paper. 

%\section*{Supplementary Materials}

\end{document}